\documentclass[%
 reprint,
 amsmath,amssymb,
 aps,
]{revtex4-2}

\usepackage{graphicx}% Include figure files
\usepackage{dcolumn}% Align table columns on decimal point
\usepackage{bm}% bold math
\usepackage[dvipsnames]{xcolor}

\begin{document}

\preprint{APS/123-QED}

\title{Collision of nanoparticles of covalently bound atoms. Impact of stress-dependent adhesion}

\author{Alexey A. Tsukanov}
\author{Nikolai V. Brilliantov}%
 \altaffiliation[Also at ]{Department of Mathematics, University of Leicester, Leicester, United Kingdom}
 \email{nb144@leicester.ac.uk}
\affiliation{Skolkovo Institute of Science and Technology, 30 Bolshoi Boulevard, Moscow, 121205, Russia
}%

\date{\today}

\begin{abstract}
The impact of nanoparticles (NPs) comprised of atoms with covalent bonding is investigated numerically and theoretically. We use recent models of covalent  bonding of carbon atoms and elaborate a numerical model of amorphous carbon (a-C) NPs, which may be applied for modelling soot particles. We compute the elastic moduli of the a-C material which agree well with the available data. We reveal an interesting novel phenomenon -- stress dependent  adhesion, which refers to stress-enhanced formation of covalent bonds between contacting surfaces. We observe that the effective adhesion coefficient linearly depends on the maximal stress between the surfaces and explain  this dependence. We compute the normal restitution coefficient for colliding NPs and explore the dependence of  the critical velocity, demarcating bouncing and aggregative collisions, on the NP radius. Using the obtained elastic and stress-dependent adhesive coefficients we develop a theory for the critical velocity. The predictions of the theory agree  very well with the simulation results.
\begin{description}
\item[Keywords]
nanoparticles, collision, aggregation, adhesion, covalent bonding, amorphous carbon, carbon black, soot, molecular dynamics
\end{description}
\end{abstract}

\maketitle

%\tableofcontents

\section{\label{sec:level1}Introduction}

In numerous natural and technological processes, collisions of micro- and nano-size particles are observed. The collision outcome generally depends on the impact velocity, material properties of particles and their structure. Small particles with not very large impact speed usually form a joint aggregate upon a collision. Solid particles suspended in the air (dust, soot, pollen), aggregate into larger one, which subside to the ground, see e.g. \cite{Drake,Hidy,Friedlander}. The concentration and size distribution of such particles in the air determines its quality. Hence it is important to develop a predictive model for particle aggregation. Aggregative collisions also play  a crucial role in many astrophysical phenomena, like planetary rings and planetesimal formation, evolution of interstellar dust clouds and other, e.g.  \cite{Saslaw,Lissauer:1993,Chokshietal:1993,DominikTilens:1997,Ossenkopf1993,SpahnAlbersetal:2004,esposito2006,BrilliantovPNAS2015}.

An important parameter which discriminates bouncing and aggregative collisions is the critical impact velocity $v_{\rm cr}$  \cite{DominikTilens:1997,SpahnAlbersetal:2004,bri2,Krijt2021}. For the impact velocities smaller than the critical one, $v \leq v_{\rm c}$, particles aggregates, while for the larger one, $v > v_{\rm c}$ --  rebound (here we do not consider the disruptive collisions).  The appearance of the critical velocity is associated with the energy losses at a collision. They are attributed to the dissipation in the bulk of viscoelastic material and to irreversible formation  of an adhesive contact at the impact. If the initial kinetic energy of the relative motion is larger than the energy losses, particles bounce, otherwise -- stick.

The critical velocity sensitively depends on many parameters, including the particles size, their bulk and surface material properties. The latter include elastic constants, viscous constants and coefficient  of adhesion \cite{bri2}. Depending on the material, different mechanisms of losses -- bulk or surface, can dominate \cite{Krijt2021}. Commonly, however, the formation of the  adhesive contact is much more energy costly then the bulk dissipation, especially for small particles.
Modelling particles collisions, e.g. \cite{BFP2018,Das,Saitoh,AnnaIcarus2012,PoeshelNeGEps}, the authors either use macroscopic interaction laws between two bodies, or microscopic inter-atomic potential for inter and intra-clusters interactions. In the former case the material parameters are assumed to be constant. In the latter one the atomic potentials, although being sometimes rather sophisticated, do not account for the chemical bonding between constituting atoms. In our study we analyze the collisions of aggregates comprised of atoms with covalent bonds. The most prominent example of such particles are carbon particles, mainly soot, -- the main contaminater of the air.

The inter-atomic interaction potential in our systems explicitly describes the formation of chemical bonds between atoms. Moreover, in the course of a contact the chemical bonds may be formed between atoms from different aggregates, that is,  the bonds may arise between clusters. Such an effect has a significant impact on the overall collision dynamics. Indeed, the formation of  the inter-cluster chemical bonds depends on the mutual compression of cluster surfaces at a contact; the compression is caused by the stress at the contact zone.  The larger the compression (and the contact  stress),  the larger the number of emerging inter-cluster bonds.  The latter noticeably contribute to the adhesion energy  of the contacting surfaces, manifesting a novel phenomenon -- stress dependent adhesion.

The contact stress at a collision depends, in its turn, on the kinetic energy of the colliding aggregates, that is, on the impact speed $v_{\rm imp}$. Hence the adhesion energy, which hinders bouncing, depends on $v_{\rm imp}$ resulting in a complicated  dependence of the critical velocity $v_{\rm cr}$ on the system parameters.  The goal of the present study is to explore and explain this dependence.
The rest of the article is organized as follows. In the next Sec. II we describe the simulation model. In Sec. III we report the simulation results. Sec. IV is devoted to  the theory of  the critical sticking velocity, for stress-dependent adhesion. Finally, in Sec. V we summarize our findings.

\section{Physical model and simulation detail}

In our study we focus on amorphous carbon (a-C) nano-particles which are the most important for applications, as the model of soot particles. The size of soot particles varies in the range from $94$ to $1000$~\AA;~ they are almost spherical in shape \cite{SHAHAD1989141}. Below we discuss in detail the model of a-C  soot particles.

\subsection{Model of bulk amorphous carbon}

To develop an adequate model  of a-C nano-particles, we start from the model of bulk material.  Using the periodic boundary conditions we apply a rapid cooling (quench) of random structure of carbon atoms. As we show below,  this approach yields the model, that reproduces rather accurately the mechanical properties of amorphous carbon -- the elastic moduli and Poisson's ratio.

Initially the carbon atoms were uniformly placed in a simple cubic lattice with the lattice constant $a = 2.152$~\AA, which provides the target density of about $2\ {\rm g/cm^3}$. After that, the atoms were randomly displaced from their initial positions in each axial direction on a distance $\delta$, uniformly distributed from $-a$ to $a$. Then the system was relaxed in periodic boundary conditions and constant volume with the energy minimization; we apply the Polak-Ribiere conjugate gradient algorithm as it is implemented in the LAMMPS software package \cite{lammps}. After the  energy minimization we performed two-stage quench: An isochoric, $V=\rm{const}$, cooling   from $T=6000~$K to $T=1700~$K for 3080~ps and then an isobaric, $p=\rm{const}$,  cooling for 1000~ps from $T=1700~$K to normal conditions with $T=300$~K and pressure $p=0.1$~MPa.

Similar to Ref. ~\cite{Orekhov2020} we used a relatively low cooling rate of $dT/dt  =\dot{T} \leq 1.4$~K/ps. It is much smaller than $\dot T = 10 - 1000$~K/ps of Ref. \cite{Jana2019}, but still larger then the experimental one of $\dot{T} \sim 10^{-4}$~K/ps ~\cite{Orekhov2020}, which is hardly achievable in present MD simulations due to a high computational cost.

As a reasonable compromise between the precision and computational expenses we used  the modified version of the Adaptive Intermolecular Reactive Empirical Bond Order potential (AIREBO) \cite{AIREBO}, with non-bonded pairwise interactions described by Morse potential (AIREBO-M) \cite{AIREBO-M}. The AIREBO-M is based on the Reactive Empirical Bond Order potential of second generation (REBO2) \cite{REBO2}. It is capable to model amorphous carbon with the density of $2.4\ {\rm g/cm^3}$ and below \cite{Li2013}. Here we  used the full version of AIREBO-M potential with the torsion term and the cutoff distance $r_{\rm cut}= 3\cdot2^{-1/6}r^{eq}_{\rm M}$, where $r^{eq}_{\rm M}$ is the equilibrium inter-atomic distance for the Morse potential \cite{AIREBO-M}. The obtained final mass density of the bulk a-C material was $\rho=2178\ {\rm kg/m^3}$. It differs from the initial density due to periodic cell relaxation at NpT conditions.

\subsection{Nanoparticle model}

\begin{figure*}
\includegraphics[width=0.85\textwidth]{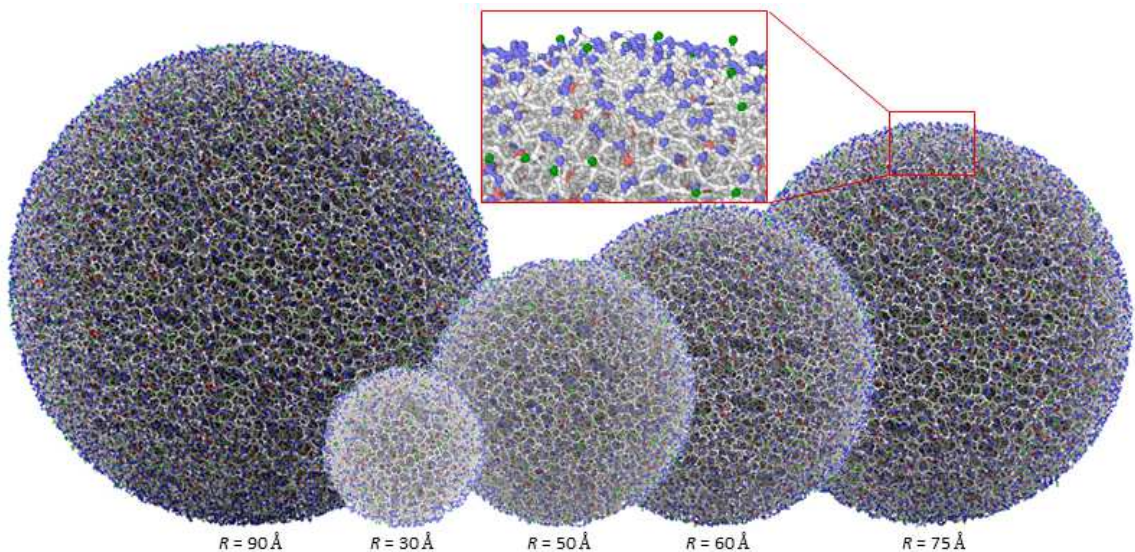}
\caption{\label{fig:nanoparticles} Five models of amorphous carbon (a-C)  nanoparticles of different size with radii 30, 50, 60, 75 and 90~$\rm\AA$. The colour coding: grey -- $sp^2$-carbon, blue -- $sp$-carbon, red -- $sp^3$-carbon and green -- hydrogen. The visualization is made with OVITO software \cite{OVITO2009,OVITO2012}.}
\end{figure*}

In order to construct nanoparticles of different size, the bulk a-C periodic cell, generated as discussed above, was periodically replicated along all three axes. Then, the spherical particles of the radii $R=30,\, 50,\, 60, \,75$ and $ 90{\rm \AA}$ were cut from the resulting volume. Nanoparticles (NPs) were relaxed at  room temperature $T=300$~K. After that, the under-coordinated carbon atoms on the surface of the NPs with the coordination number $n_{\rm coor}=1$ (``chemical radicals'') were replaced by the hydrogen (H) atoms; the valence rule is obviously satisfied, as the valence of H atom is 1. To find  the coordination number, we computed  the radial distribution function (RDF) for C-C pairs,  see Fig. \ref{fig:rdf} of  the Appendix~A, and used the characteristic distance of $r^{\rm b}_{1-2}\approx 1.85$~$\rm\AA$. We considered two atoms as bounded if their inter-center distance does not exceed   $r^{\rm b}_{1-2}$. This value of $r^{\rm b}_{1-2}$ corresponds to the maximal extension of the first peak of the RDF, see Fig. \ref{fig:rdf}. Atoms with two neighbours,  $n_{\rm coor}=2$, are  considered as $sp$-hybridized (they may form a linear polymer), with three neighbours, $n_{\rm coor}=3$, -- as $sp^2$-hybridized (may form a planar compound), and with the four ones, $n_{\rm coor}=4$, -- $sp^3$-hybridized (may form a tetrahedral structure). Note that our model lacks over-coordinated C atoms with $n_{\rm coor}\geqslant5$.
The constructed nanoparticles are depicted in Fig.~\ref{fig:nanoparticles} and the parameters of the model are summarized in Table~\ref{tab:table1}.

\begin{table}[b]
\caption{\label{tab:table1}
Parameters of a-C nanoparticles models.}
\begin{ruledtabular}
\begin{tabular}{cccccc}
 $R$, \AA & $n_{\rm at}$\footnotemark[1] & ${sp}$-C, \% & ${sp^3}$-C, \% & H, \%& $[{\rm H]_s}$\footnotemark[2]\\
\hline
30 & 12.4 & 9.7 & 2.9 & 1.0 & 1.11 \\
50 & 57.2 & 6.9 & 3.1 & 0.6 & 1.09 \\
60 & 100 & 6.2 & 3.1 & 0.5 & 1.06 \\
75 & 193 & 5.5 & 3.2 & 0.4 & 1.14 \\
90 & 334 & 5.0 & 3.2 & 0.4 & 1.24 \\
\end{tabular}
\end{ruledtabular}
\footnotetext[1]{Total number of atoms in the NP, $\times 10^3$.}
\footnotetext[2]{Surface concentration of H atoms, in ${\rm nm^{-2}}$.}
\end{table}

\section{Simulation results}

As it follows from the previous studies of particles collisions, the crucial role in the impact dynamics play such physical properties as  elastic moduli, Poisson's ratio and the adhesion coefficient of particles surface \cite{BFP2018,Saitoh,AnnaIcarus2012,PoeshelNeGEps,bri2,Krijt2021,BrilliantovSpahn2006}. Therefore it is worth to measure these quantities for our model of a-C.

\subsection{Elastic moduli and Poisson's ratio}

The values of Young's modulus $Y$ and Poisson's ratio $\nu$ for the model material can be obtained by means of non-equilibrium molecular dynamics (NEMD) simulations, namely, by MD of slow (quasi-static) uniaxial expansion and the stress-strain curve analysis. By this way $Y$ may  be evaluated from  the slope of the stress-strain curve, and $\nu$ can be obtained directly as a ratio between the transverse contraction and axial (longitudinal) strain \cite{Landau:1965}. However, in order to obtain accurate estimates we performed two series of independent simulations: (a) Three simulations of the uniaxial extension along each of the three axis independently at constant temperature $T=300$~K and load-free conditions for other four sides. That is, if the extension was  along $x$-axis, the normal stresses at the boundaries $y=0$, $y=y_{max}$, $z=0$ and $z=z_{max}$ was kept zero, $\sigma_{yy}=\sigma_{zz}=0$); this  yielded the Young modulus. (b) Two simulations of uniform isothermal compression-expansion to evaluate the bulk modulus $K$. Then the Poisson's ratio can be found from the following expression,
\begin{equation}
\nu =  \frac{1}{2}\left(1-\frac{Y}{3K}\right)%
\label{eq:nu}.
\end{equation}
In both series of simulations the slow enough strain rate was used, $\dot\varepsilon=10^7\, {\rm s^{-1}}$, which was a few order of magnitude smaller that the strain-rate of $10^8 - 10^{10}\, {\rm s^{-1}}$ exploited in Ref. \cite{Hossain}. Using the regime of linear deformations, we obtained the following results for the elastic moduli (see Appendix A for more detail):

\begin{equation}
\label{YK}
Y = 269.2\pm 27.1\ {\rm GPa,}\quad K = 191.9\pm 12.6\ {\rm GPa.}
\end{equation}
Substituting the above values into the Eq.~(\ref{eq:nu}) yields the Poisson's ratio, $\nu = 0.266$.

It is interesting to compare the obtained quantities with the available in literature data for similar materials. Our results are in a very good agreement with the estimates obtained in Ref. \cite{Jana2019} by the density functional theory.  Indeed, for  the a-C with density $\rho \simeq  2125 - 2325\ {\rm kg/m^3}$ the authors reported $Y \approx 250 - 320\ {\rm GPa}$, $K \approx 175 - 220\ {\rm GPa}$ and $\nu \approx 0.25$, which  are very close to the above result, Eq. \eqref{YK}. This strongly supports our choice of the AIREBO-M potential for an adequate modelling of amorphous carbon; further justification for the usage of this potential is presented in Appendix A.

With the obtained estimates for $Y$, $\nu$ and $\rho$ one can find the longitudinal $c_l$ and transverse $c_t$ sound velocities in the bulk of the material,
$$c_l = \left[\frac{Y(1-\nu)}{\rho(1+\nu)(1-2\nu)}\right]^{1/2}; \qquad c_t = \left[\frac{Y}{2\rho(1+\nu)}\right]^{1/2},  $$
giving $c_l \simeq 12380\ {\rm m/s}$ and $c_t  \simeq  6570\ {\rm m/s}$. These estimates allow to assess the collision regime -- super-sonic, subsonic, or  quasi-static  for different impact velocities \cite{BrilliantovSpahnHertzschPoeschel:1994}.

\subsection{Stress dependent adhesion}

The adhesion characterizes the strength of the inter-surface contact. It is quantified by the work $W_{\rm adh}$ needed to move the surfaces from their contact  to the infinite separation; it is done against the attractive surface forces. The specific work per unit area is called the adhesion coefficient $\gamma$. For the contact of the same material $\gamma$ is twice the surface tension of this material in vacuum \cite{Israelachvilli2011}.

\begin{figure*}
\includegraphics[width=1.0\textwidth]{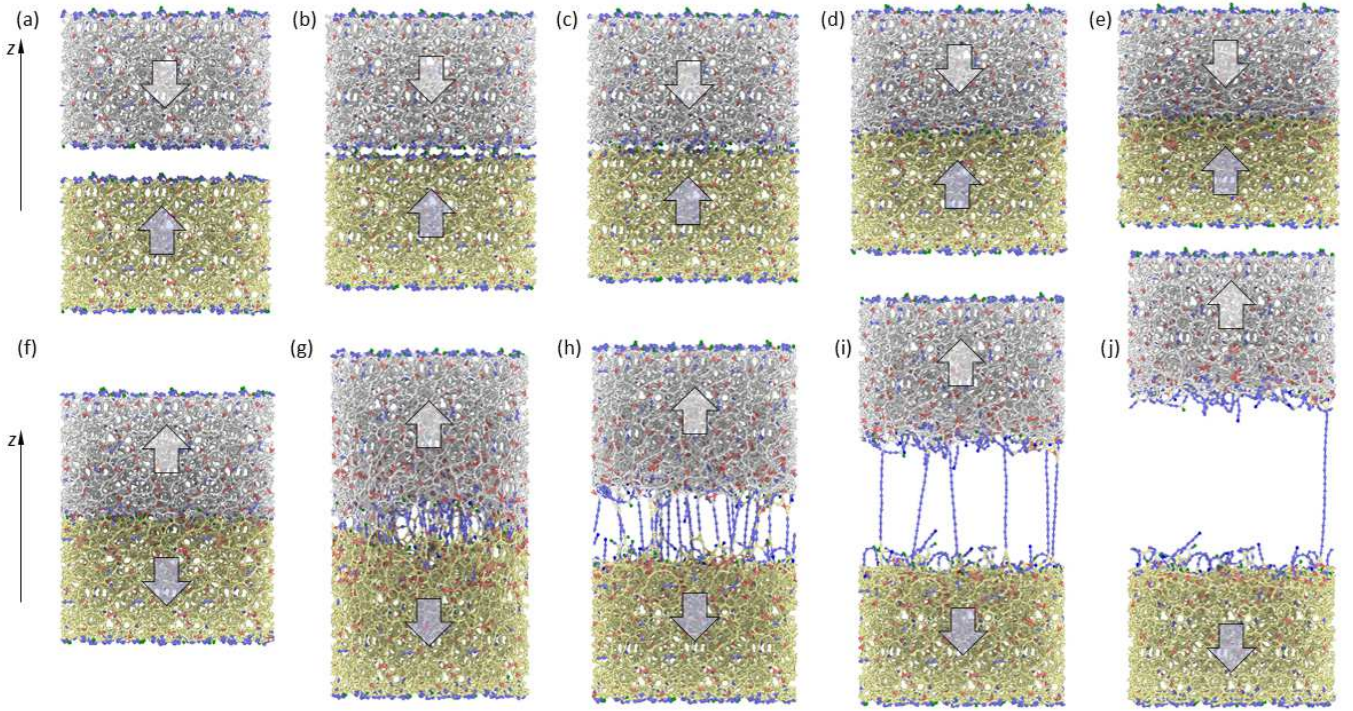}
\caption{\label{fig:smd3d} Frame sequences of forward (a-e) and reverse (f-j) cv-SMD simulations. External forces $\pm F_{\rm z}^{\rm ext}$ (light-blue arrows) are  applied to each atom of the infinite a-C blocks to enforce the adhesion (the forward process) or to split the system back into two parts (the reverse process). The periodic boundary conditions are  applied along $x$ and $y$ axis. The frames are: (a) close to the initial configuration, (b) the first bonds are formed between the most protrudent points of the surfaces; the  stress is  $\sigma=1.13$~GPa, (c) adhesion with the bond formation over the entire contact area ($\sigma=8.52$~GPa), (d, e) -- further loading up to stress 34.84 and 63.65~GPa, respectively; (f) -- the beginning of the reverse process, (g) first bonds are broken, (h) more than  half of the bonds are broken, and (i, j) the last few bonds, formed by linear carbon chains (carbyne), still exist;  the colour coding is the same as in  Fig.~\ref{fig:nanoparticles}, except for that $sp^2$-C atoms of the second plate are yellow.}
\end{figure*}

The parameter $\gamma$ may be estimated with the use of the bidirectional constant-velocity steered MD (cv-SMD) simulation with the potential of mean force (PMF) \cite{Izrailev}. To apply this technique, one needs to prepare a model of two a-C plates with hydrogenated surfaces, infinite in $x$ and $y$ directions, see Fig.~\ref{fig:smd3d}a. We used the computational box with the dimensions of $50.618 \times 50.134 \times 200~\rm \AA$ and periodic boundary conditions along $x$ and $y$ axis. The surface concentration of hydrogen atoms was  $1.10\ nm^{-2}$.

\begin{figure*}
\includegraphics[width=1.0\textwidth]{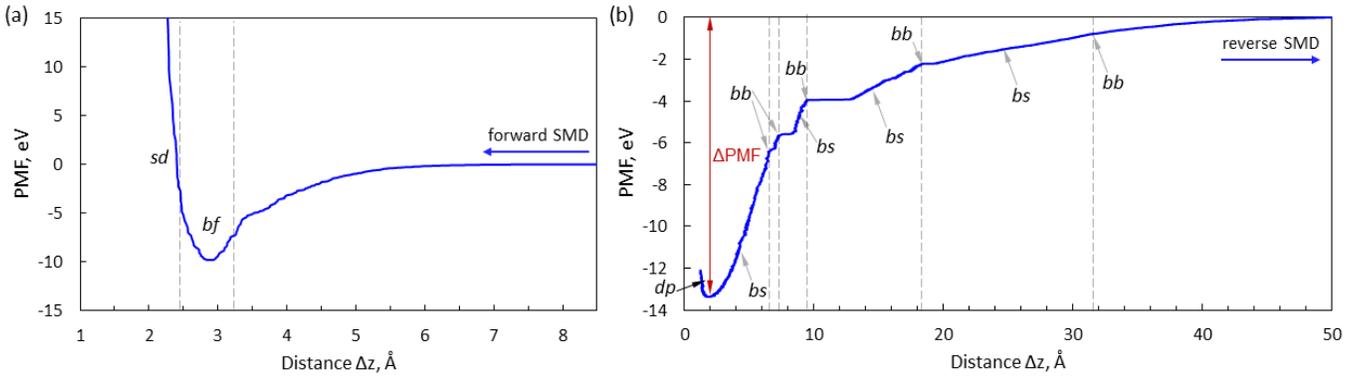}%{FIG_PMF_DRAFT.eps}
\caption{\label{fig:pmf} The potential of the mean force  ${\rm PMF}(\Delta z)$ for the forward, (a), and reverse, (b),  SMD simulations, as the function of inter-surface distance $\Delta z$.  The forward process refers to the enforced adhesion, the reverse -- to the surface separation.
During the forward process (panel a) ${\rm PMF}(\Delta z)$ decreases with the decreasing $\Delta z$ and becomes negative due to van der Waals attraction.  Further decrease of the distance is accompanied by the formation of inter-surface covalent bonds ($bf$ in the figure), reflected in the rapid decrease of ${\rm PMF}(\Delta z)$ towards the minimum at  $\Delta z \approx 3\ \rm\AA$. For still smaller $\Delta z$ the material undergoes a strong deformation ($sd$) with the steep increase of the ${\rm PMF}(\Delta z)$.  At the beginning of the reverse process (panel b) the plates become unloaded from the previous compression, undergoing the depression ($dp$).  Then the multiple stages of bonds stretching ($bs$) and bonds breaking ($bb$) are repeatedly observed.  The total difference of  $\Delta{\rm PMF}$ in the reverse process --  between the point of the unloaded (relaxed) contact and the point, where plates are completely separated, gives the estimate of the work of adhesion (red arrow).}
\end{figure*}

The bidirectional simulation includes two stages -- the forward and reverse one. In the former stage the applied external force pulls the surfaces of the plates into a contact and then press them together; this causes an enforced adhesion, see Fig.~\ref{fig:smd3d}a-c.
The later stage corresponds to the forced separation of the surfaces. In this case the external force performs the work to separate the plates, breaking all covalent bonds emerging during the forced adhesion, see Fig.~\ref{fig:smd3d}d-f.   Measuring this work as a function of the distance between surfaces at large separation in the reverse  stage, one can estimate the adhesion energy. The dependence of the work of the external force, which is also called the potential of mean force (PMF), on the surface separation $\Delta z$ is depicted in Fig. \ref{fig:pmf}, yielding the estimate,  $W_{\rm adh} \simeq |\Delta{\rm PMF}|$.

\begin{figure*}
\includegraphics[width=0.9\textwidth]{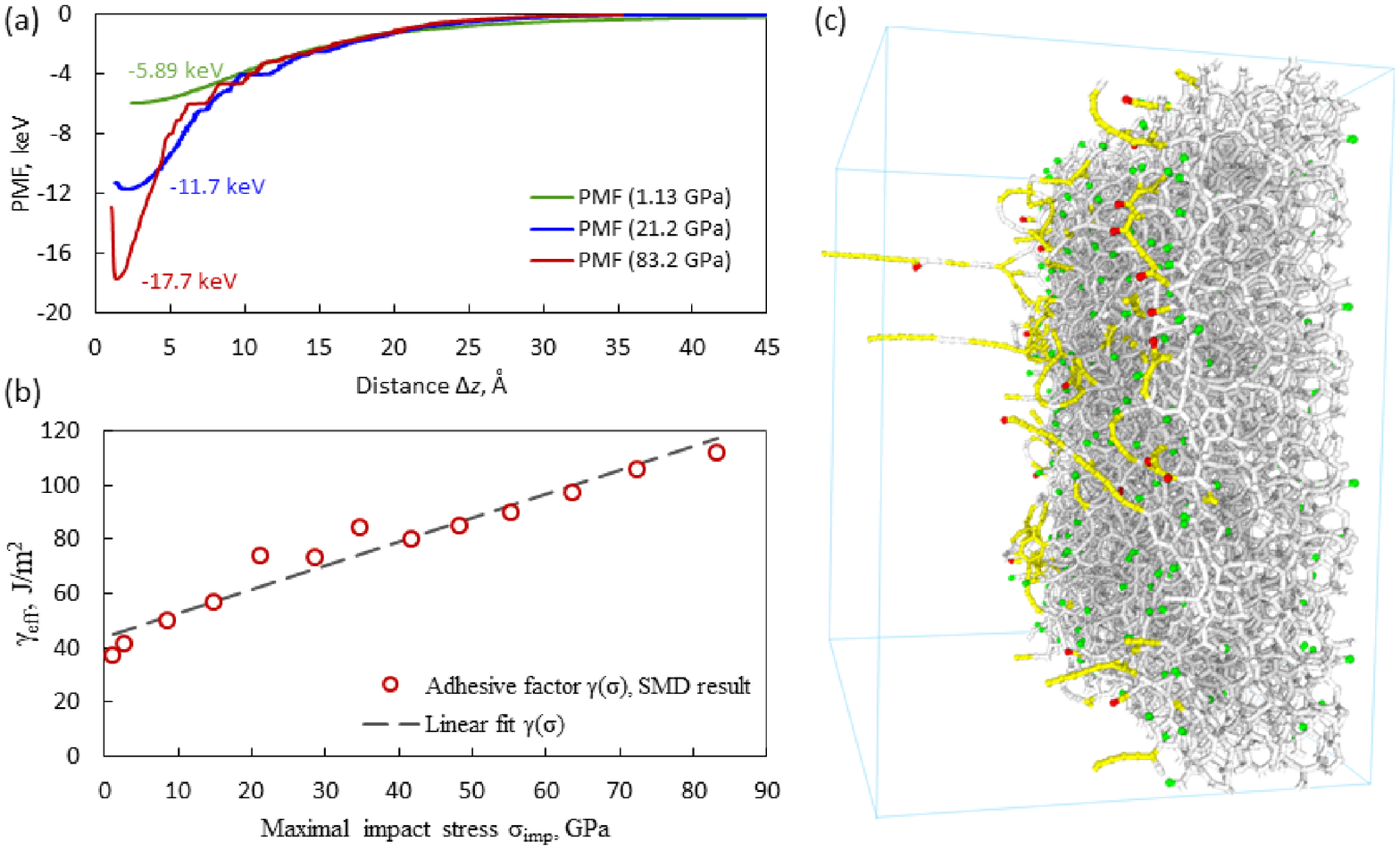}
\caption{\label{fig:pmf3} The results of cv-SMD simulations: (a) The potential of the mean force between surfaces as the function of their  separation $\Delta z$ after a contact with different maximal impact stress $\sigma_{\rm max} \simeq 1.13$ (green), 21.2 (blue) and 83.2~GPa (red). (b) The adhesion coefficient $\gamma$ as the function of the maximal stress, $\gamma = \gamma(\sigma_{\rm max})$. Points -- the simulation data, from 13 independent runs; dashed line -- the linear fit for $\gamma = \gamma(\sigma_{\rm max})$; (c) The structure of the surface after the compression and complete separation of a-C plates. The colour coding: grey and green -- C and H atoms, that initially belonged to this plate; yellow and red -- C and H atoms, that initially belonged to the different plate.}
\end{figure*}

The results of cv-SMD simulations are illustrated in Fig.~\ref{fig:smd3d} and Fig.~\ref{fig:pmf}.  During the first stage of SMD, the plates move form their initial positions,  toward each other under the action of the external force,   Fig.~\ref{fig:smd3d}a. The force is modelled by a harmonic spring connecting the centers of mass of two plates, with a uniformly contracting equilibrium  length $l_{\rm eq}$. The force is distributed among all atoms of the plates, proportionally to the masses of atoms, and the equilibrium length decreases with the constant  rate of $\dot{l}_{\rm eq}=0.02~ {\rm \AA/ps}$, which is slow enough to guarantee the quasi-static process.  At the moment of contact, covalent C-C bonds are formed between the surfaces, Fig.~\ref{fig:smd3d}b. With the further approach, the elastic stress $\sigma_{zz}$ increases and the plates undergo deformation. At this moment new covalent bonds are formed Fig.~\ref{fig:smd3d}c. Note, that zero-stress conditions were applied at the periodic boundaries $\sigma_{xx}=\sigma_{yy}=0$. After achieving the maximal stress, the contraction of the equilibrium length of the virtual spring altered to  a uniform elongation, with the same rate of $0.02~{\rm \AA/ps}$. This leaded to the  separation of contact surfaces Fig.~\ref{fig:smd3d}d. During the separation, linear chains of $sp$-hybridized carbon -- the carbynes \cite{Carbyne}, were  formed between the surfaces, Fig.~\ref{fig:smd3d}e,f. Given that the carbyne has a high tensile strength, these chains made the separation of the plates much more difficult. The interacting surfaces underwent noticeable structural changes. For instance,  after the complete separation, a surface of one of the plates demonstrated loops and chains, formed by $sp$-C atoms from another plate, see  Fig.~\ref{fig:pmf3}, where the yellow carbyne chains are comprised by foreign atoms (Fig.~\ref{fig:pmf3}c). As it can be seen, after the separation the number of $sp$-hybridized carbon atoms increased.

Hence the adhesion-separation is a strongly irreversible process, with the formation and breakage of  covalent bonds. The  absolute value of the external force work is much larger in the reverse process then in the forward one. The excess energy of this irreversible process is adsorbed by the thermostat. For the setup depicted in Fig.~\ref{fig:pmf}, the maximal  stress was $\sigma_{\rm max}=35$~GPa and $W_{\rm adh} \simeq |\Delta{\rm PMF}|= 13.35$~keV for the surface area  of $S = 50.618 \times 50.134\ {\rm \AA}^2$. This yields the adhesion coefficient $\gamma= W_{\rm adh}/S=84.3\, \rm{J/m^2}$. With the larger maximal stress $\sigma_{\rm max}$ the larger number of new bonds can be formed. Indeed, owing to the deformation of the material, a larger number of atoms from different surfaces are brought into a contact with  each other. If the deformation is strong enough, even atoms from the subsurface layers  can form bonds with the atoms from another plate.
Next,  we explored how the separation work $|\Delta{\rm PMF}|$ depends on the maximal contact  stress $\sigma_{\rm max}$. We performed a series of additional cv-SMD simulations for  different $\sigma_{\rm max}$. The according simulation results are presented in Fig.~\ref{fig:pmf3}(a, b).

As it can be seen $\gamma(\sigma_{\rm max})$ monotonically increases with increasing maximal contact stress,  Fig.~\ref{fig:pmf3}b. Qualitatively, this may be understood as follows. With the increasing stress, two surfaces come closer and closer to each other, which results in the increasing number of atoms that can form covalent bonds. All these bonds become broken when the surfaces are separated. The energy needed to break all the emergent bonds contributes to the adhesion energy and thus to the adhesion coefficient. Generally, one can write for the adhesion coefficient:
\begin{equation}
\gamma = \gamma_{\rm non-b} + \gamma_{\rm bond} =\gamma_{\rm non-b}+n_{\rm s.c.} \left< E_{\rm b} \right> ,
\label{eq:gengam}
\end{equation}
where $ \gamma_{\rm non-b}$ characterizes the part of the adhesion coefficient, associated with the non-bonding interactions, while $\gamma_{\rm bond}$ -- with the broken bonds emerging at a contact. The latter may be written as a product of the surface density of emerging bonds, which depends on the maximum  stress, $n_{\rm s.c.} = n_{\rm s.c.} (\sigma_{\rm max})$ and the average energy per the bond, $\left< E_{\rm b} \right>$. Naturally, the surface density of the emergent bonds is proportional to the number  of surface atom of the different  plates in a tight contact. The later increases with the compression $|\Delta z - d_{\rm eq}|$ and thus with the contact stress $\sigma_{\rm max}$. Except for the initial part, where the surface roughness dominates, the dependence $n_{\rm s.c.} (\sigma_{\rm max})$ is almost linear, that is $n_{\rm s.c.}= n_{\rm s.c.0} + a \sigma_{\rm max}$, where $n_{\rm s.c.0}=n_{\rm s.c.} (\sigma=0)$ is the number of surface contacts for vanishing contact stress, see Fig. \ref{fig:contacts} of the Appendix B.  $\left< E_{\rm b} \right>$ may be also approximated by a constant, which yields, the linear dependence for $\gamma (\sigma_{\rm max})$:

\begin{equation}
\gamma = \gamma_0 + \varkappa \sigma_{\rm max},
\label{eq:gammaeff}
\end{equation}
where  $\gamma_0=\gamma_{\rm non-b} + n_{\rm s.c.0}  \left< E_{\rm b} \right>$ and $\varkappa =a \left< E_{\rm b} \right>$. The least squares fit of the data depicted in Fig. \ref{eq:gammaeff} yields
$$ \gamma_0 = 44.775\ {\rm J/m^2}, \qquad \quad \varkappa = 0.8503\cdot10^{-9}\ {\rm m}.
$$
Knowing the dependencies $\gamma (\sigma_{\rm max})$ and  $n_{\rm s.c.}(\sigma_{\rm max})$, shown in Fig.  \ref{fig:contacts}, one can estimate
the average bond energy, $\left< E_{\rm b} \right> \simeq 14\cdot 10^{-18}$J. This is about 1.5 times larger than the energy of $9.6\cdot 10^{-18}$J, needed to break a carbyne chain bond, see Appendix B for detail.

\subsection{Coefficient of normal restitution and critical velocity}

\begin{figure*}
\includegraphics[width=1.0\textwidth]{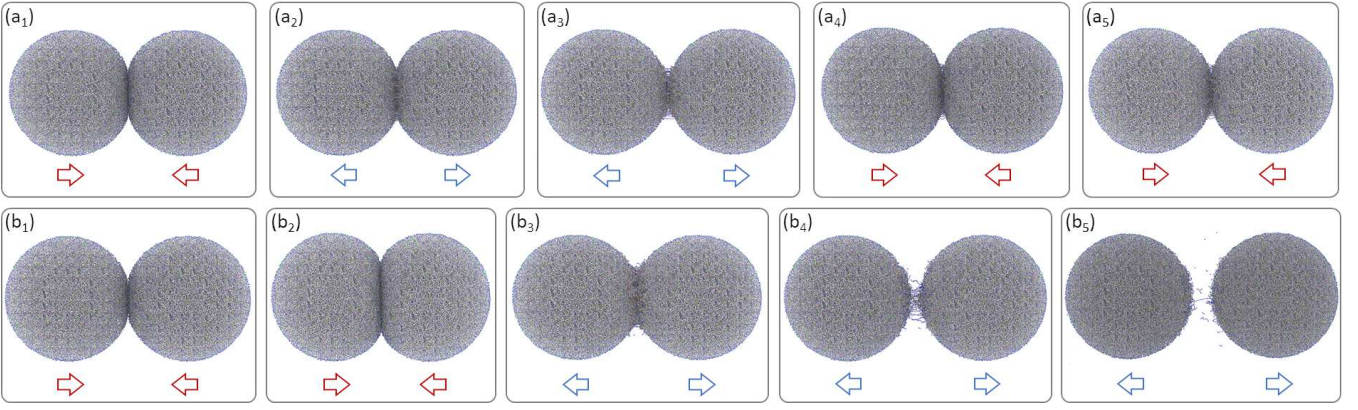}% Here is how to import EPS art
\caption{\label{fig:collision} The collision of two nanoparticles of radii $R=90\ {\rm\AA}$. The upper panel  illustrates the subsequent frames for the  collision with the impact velocity $v=1000$~m/s, which is  smaller than the critical one, $v_{\rm cr}$. The bottom panel -- for the collision with the impact velocity $v=2000$~m/s, exceeding the critical velocity $v_{\rm cr}$.}
\end{figure*}

We performed a series of about forty simulations of head-on collisions, varying the impact velocity $v_{\rm imp}$ and nanoparticles  size $R$. We used classical molecular dynamics (MD) simulations with the microcanonical ensemble, that is, we explored adiabatic collisions, where the total energy of the system is conserved. The set of  impact snapshots for two colliding NPs of radii $R=90\ \rm\AA$ is presented in Fig.~\ref{fig:collision}. The snapshots (${\rm a}_i$) on the upper panel illustrate the aggregative collision for the impact velocity 1000~m/s; the covalently bonded aggregate is formed. On the contrary, the snapshots (${\rm b}_j$) on the bottom  panel show the bouncing collision with the impact velocity 2000~m/s.

The after-collision relative velocity $v_{\rm fin}$ of the particles was measured for each impact, and the coefficient of normal restitution $\epsilon$ was evaluated, according to its definition, e.g.  \cite{BrilliantovPoeschelOUP},
$$
\epsilon = \frac{v_{\rm fin}}{v_{\rm imp}},
$$
as the  function of the impact velocity $\epsilon=\epsilon(v_{\rm imp})$.

\begin{figure*}
\includegraphics[width=1.0\textwidth]{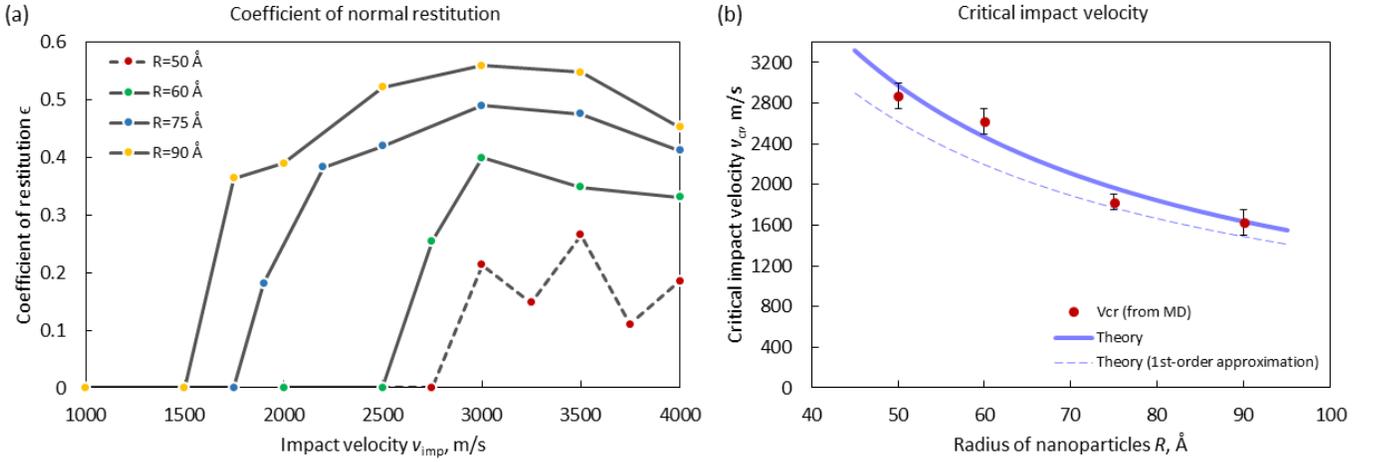}% Here is how to import EPS art
\caption{\label{fig:restitution} (a) The coefficient of normal restitution $\epsilon$ as a function of the impact velocity $v_{\rm imp}$ for different radii $R$ of a-C nanoparticles. (b) The dependence  of the critical velocity $v_{\rm cr}$ on the nanoparticle radius $R$. Dots -- simulation results;  thick blue line -- theory, the solution of Eq. \eqref{eq:vcvc}; dashed blue line -- theory, first-order approximation Eq. \eqref{eq:vford}.
}
\end{figure*}

The results of these simulations for $R =$ 50, 60, 75 and 90~$\rm\AA$ are presented in Fig.~\ref{fig:restitution}a. As it can be seen from the figure, for the impact velocity below some threshold $v_{\rm cr}$, the coefficient of normal restitution becomes zero,  $\epsilon=0$; this means that the particles stick together. Moreover, in accordance with the theory \cite{bri2}, the critical sticking velocity for  $v_{\rm cr}$ decreases with the increasing particle size. Fig.~\ref{fig:restitution}b (points) illustrates the obtained dependence of the critical velocity on the NP radius, $v_{\rm cr}=v_{\rm cr}(R)$.

\section{Theory}
The theory of the critical sticking velocity has been developed in Ref.\cite{bri2} for particles with usual (stress-independent) surface properties. Here we generalize it for the surfaces that can form covalent bonds and thus demonstrate stress-dependent adhesion. Generally, the energy conservation in a head-on collision reads,

\begin{equation}
\frac{m_{\rm eff}v_{\rm imp}^2}{2} = \frac{m_{\rm eff}v_{\rm fin}^2}{2} + W_{\rm adh} + W_{\rm vis},
\label{eq:energy}
\end{equation}
where $m_{\rm eff}=\frac{m_1m_2}{m_1+m_2}=\frac12 m$ is the effective mass ($m=m_1=m_2$  is the particle mass). Eq. \eqref{eq:energy} simply states that the initial kinetic energy of the relative motion  is equal to the final kinetic energy plus the energy losses due to the viscous dissipation in the bulk of  the material, $W_{\rm vis}$ and irreversible work of adhesion $W_{\rm adh}$, associated with the formation and breakage of the contact at the impact. For the case of sticking velocity $v_{\rm fin}=0$. Commonly, the work of adhesion for small particles significantly exceeds the viscous losses and hence $W_{\rm vis}$ may be neglected. For particles which form the surface covalent bonds, the role of the viscous term is even less important, which leads to the estimate of the critical velocity \cite{bri2}:
\begin{equation}
v_{\rm cr} =\sqrt{\frac{2 W_{\rm adh}}{m_{\rm eff}}}.
\label{eq:vcr}
\end{equation}
The adhesion work $W_{\rm adh}$ has been derived for the JKR interaction law \cite{JKR:1971} and reads  \cite{bri2},

\begin{equation}
W_{\rm adh} = q\left(\pi^5 \gamma^5 R_{\rm eff}^4D^2\right)^{1/3}
\label{eq:W},
\end{equation}
where $R_{\rm eff}=\frac{R_1R_2}{R_1+R_2}=\frac12 R$ is the effective radius, $D=(3/2) (1-\nu^2)/Y$ is the elastic constant and $q=1.57$ is the pure number.

The above expression for $W_{\rm adh}$, has been obtained with the JKR theory, based on the continuous macroscopic  description of particles. It uses the bulk material constants,  $Y$ and $\nu$ and the adhesion coefficient  $\gamma$. Moreover JKR is a quasi-static theory. Following Ref. \cite{Saitoh}, where the JKR theory has been successfully applied for colliding NPs, we assume that Eqs. \eqref{eq:vcr} and \eqref{eq:W} remain valid for our system. Indeed, for the impact velocities equal to $v_{\rm cr}$ the particles do not separate, Fig.\ref{fig:collision}a, so that the carbyne chains, Fig. \ref{fig:smd3d}g-j, that make the difference with the common adsorption, are not formed. We also consider the impact velocities considerably smaller than the speed of sound in the material, which allows to exploit for the collisions the quasi-static approximation  \cite{BrilliantovSpahnHertzschPoeschel:1994}.

Still, the main difference between  Eq. \eqref{eq:vcr} and the conventional theory for $v_{\rm cr}$  \cite{bri2} is the stress dependence of the adhesion coefficient  $\gamma$, which implies its dependence on the impact velocity. In other words, the larger the impact velocity, the larger the compression of particles at a collision and hence, the larger the contact stress, implying  larger $\gamma$ and $W_{\rm adh}$. As the result, $W_{\rm adh}=W_{\rm adh}(v_{\rm cr})$, and \eqref{eq:vcr} is not anymore an   explicit expression for $v_{\rm cr}$,  but the equation for this quantity which is to be solved.

Since we know the dependence of $\gamma$ on the maximal contact stress $\sigma_{\rm max}$, Eq. \eqref{eq:gammaeff}, we need to find the dependence of $\sigma_{\rm max}$ on the impact velocity. Let $F_{\rm max}$ be the maximal  force at a collision of two particles, corresponding to  the  maximal contact radius $a_{\rm max}$, then
\begin{equation}
\sigma_{\rm max}  =\frac{F_{\rm max}}{\pi a_{\rm max}^2}
\label{eq:sigma},
\end{equation}
To estimate the above quantities we ignore the adhesion and dissipation and apply the Hertz contact theory, see e.g. \cite{Landau:1965}, which expresses the force and the contact radius as the function of the particles compression,  $\xi\equiv R_1+R_2-|{\bf r}_{12}|=2R-|{\bf r}_{12}|$, where ${\bf r}_{12}= {\bf r}_2-{\bf r}_1$ is the inter-center vector:
\begin{equation}
F=\frac{\sqrt{R_{\rm eff}}}{D}\xi^{3/2}; \qquad \quad a=\sqrt{R_{\rm eff} \xi}.
\label{eq:Fa}
\end{equation}
Application of Eq. \eqref{eq:Fa} to the maximal compression, $\xi=\xi_{\rm max}$, yields,

\begin{equation}
\sigma_{\rm max}  =\frac{1}{\pi D}\left(\frac{2\xi_{\rm max}}{R}\right)^{1/2},
\label{eq:sigmaH}
\end{equation}
where we use $R_{\rm eff}= R/2$. To estimate $\xi_{\rm max}$ we again apply the Hertz theory, which expresses the elastic energy of compression at a collision as $U_{\rm el}=(2/5)R_{\rm eff}^{1/2}\xi^{5/2}/D $. Neglecting again the viscous losses and the work of adhesion, we obtain:
\begin{equation}
\frac{m_{\rm eff}v_{\rm imp}^2}{2}\approx U_{\rm el} (\xi_{\rm max}) = \frac25 \frac{\sqrt{R_{\rm eff}}}{D}\xi_{\rm max}^{5/2}.
\label{eq:energyH}
\end{equation}
Combining Eqs. (\ref{eq:sigmaH}) and (\ref{eq:energyH}), and using $m_{\rm eff} = \frac12 m = \frac{2}{3}\pi\rho R^3,$ we find,
\begin{equation}
\sigma_{\rm max} = \left(\frac{20 \rho }{3  \pi^4 D^4} \right)^{1/5} v_{\rm imp}^{2/5}.
\label{eq:sigma_Vimp}
\end{equation}
Substituting $\gamma (\sigma_{\rm max})$ given by Eq. (\ref{eq:gammaeff}) into Eq. (\ref{eq:W}),  with $\sigma_{\rm max}$ from Eq. \eqref{eq:sigma_Vimp} we obtain $W_{\rm adh}$ as the  function of $v_{\rm imp}$. Substituting then the obtained expression with $v_{\rm imp} =v_{\rm cr}$ into Eq. \eqref{eq:vcr} we arrive at,
\begin{equation}
v_{\rm cr} =v_{\rm cr.0}\left(1 +B\, v_{\rm cr}^{2/5} \right)^{5/6},
\label{eq:vcvc}
\end{equation}
where
\begin{eqnarray}
v_{\rm cr.0} &=& \left( \frac{3q\pi^{2/3}}{2^{4/3}}\right)^{1/2} \frac{D^{1/3} \gamma_0^{5/6}}{\rho^{1/2} R^{5/6}}
\label{eq:v0vcr}
\\  B &=&  \left( \frac{20 \rho}{3 \pi^4 D^4} \right)^{1/5} \frac{\varkappa}{\gamma_0}.
\label{eq:Bvcr}
\end{eqnarray}
Zero-order approximation $v_{\rm cr} =v_{\rm cr.0}$ (stress-independent $\gamma$) may be substituted into Eq. \eqref{eq:vcvc}, yielding the first-order approximation:
\begin{equation}
v_{\rm cr} =v_{\rm cr.0}\left(1 +B\, v_{\rm cr.0}^{2/5} \right)^{5/6},
\label{eq:vford}
\end{equation}
with $v_{\rm cr.0}$ from Eq. \eqref{eq:v0vcr}. Generally, the  solution of the transcendental equation \eqref{eq:vcvc} may be found numerically.  The theoretical results for $v_{\rm cr}$ are compared in Fig.~\ref{fig:restitution}b with the results of MD simulations. As it can be seen from the figure the theoretical  predictions for the critical sticking velocity agree very well with the MD results. The results for the smallest nanoparticles of radii $R=30\ \rm\AA$ are not shown in Fig.~\ref{fig:restitution}b, as we did not detect bouncing collisions for such particles.

\section{Conclusions}

We explore numerically and theoretically the collision of nanoparticles (NPs) comprised of atoms with covalent bonds. We use  the AIREBO-M potential that describes covalent interaction of carbon atoms and analyse the models of amorphous carbon (a-C). The a-C nanoparticles may serve as a realistic  model of soot particles. Therefore the knowledge of the collision dynamics of such particles is important for many practical applications. Using MD simulations we investigate the bulk properties of a-C material and compute its elastic coefficients -- Young modulus and Poisson ratio. Next, we analyse the surface contact properties and reveal a novel phenomenon -- the stress dependent adsorption. It is related to the stress-enhanced formation  of covalent bonds between the contacting surfaces: The larger the stress, pressing the surfaces together, the larger the number of newly formed covalent bonds. Hence the larger work is needed to separate the surfaces. This results  in the adhesion coefficient $\gamma$ (the work against the adhesion per unit surface) that depends on the maximal normal stress between two surfaces $\sigma_{\rm max}$. We observe that  $\gamma(\sigma_{\rm max})$ may be approximated by a linear dependence almost in all studied range of $\sigma_{\rm max}$.

We analyse the dependence of the restitution coefficient of colliding particles on the particles size and impact velocity. Also we explore  the dependence of the critical velocity $v_{\rm cr}$ (the velocity demarcating bouncing and sticking collisions) on the particle radius. We develop a theory of collision dynamics of adhesive particles, with stress-dependent adhesion, and  find the dependence of  $v_{\rm cr}$ on the particle radius. Our theoretical predictions for the critical velocity are in a very good agreement with the  molecular dynamic results. We believe that the proposed approach, tested for nano-particles, will allow to model the collision dynamics of particles comprised of atoms with covalent bonds (including soot particles),  in  their all size range.

\begin{acknowledgments}

The study was supported by a grant from the Russian Science Foundation No.~21-11-00363, https://rscf.ru/project/21-11-00363/.

The research is carried out using the resources of ``Zhores'' supercomputer of Center for Computational and Data-Intensive Science and Engineering (CDISE), Skolkovo Institute of Science and Technology (Skoltech) \cite{Zhores} and the equipment of the shared research facilities of HPC computing resources of Lomonosov Moscow State University \cite{Voevodin2019, Adinets2012}.
\end{acknowledgments}

%\newpage
\appendix

\section{Structural and elastic properties of a-C}

\begin{figure}[b]
\includegraphics[width=1.0\columnwidth]{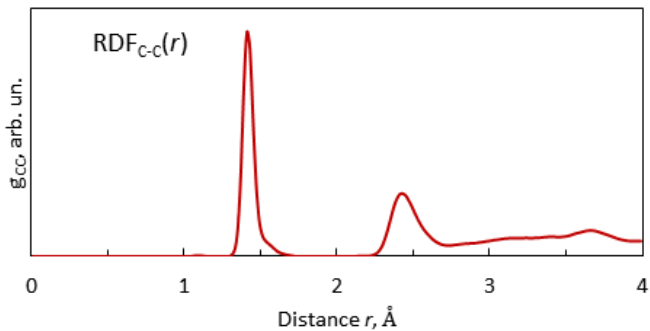}
\caption{\label{fig:rdf}The radial distribution function ${\rm RDF}_{\rm C-C}(r)$ for $C-C$ atoms. }
\end{figure}

The radial distribution function (RDF) for C-C pairs in  a-C nanoparticles is shown in Fig.~\ref{fig:rdf}. The ${\rm RDF}_{\rm C-C}(r)$ was obtained for NPs of the radius $R=90\ {\rm\AA}$ at  temperature $T=300$~K before collision. The RDF was then used to compute the distribution of the number of covalent bonds between the atoms (see the main text).

\begin{figure}[!]
\includegraphics[width=1.0\columnwidth]{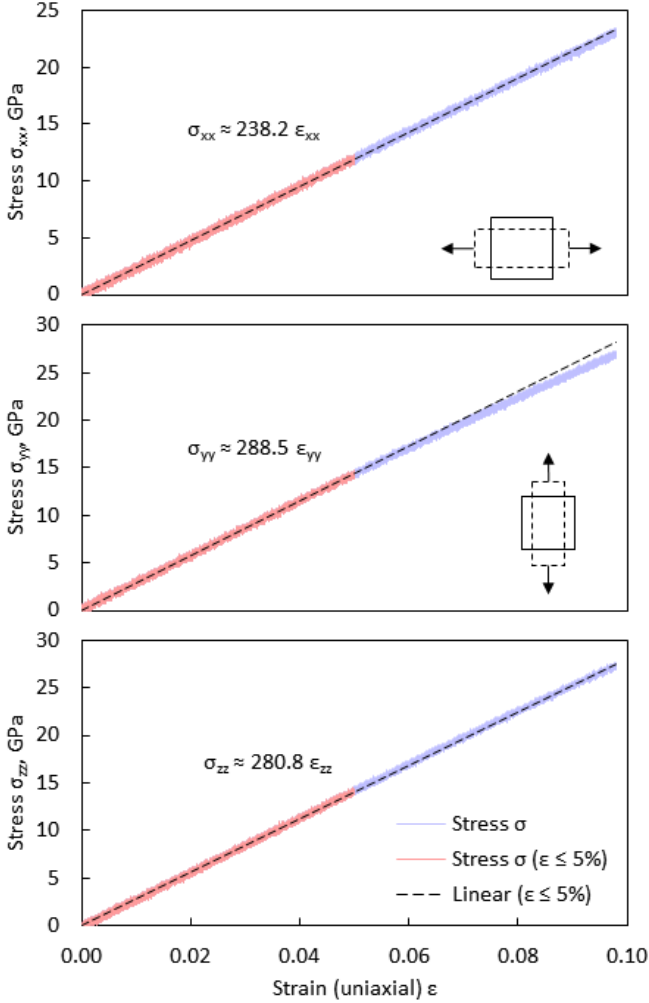}
\caption{\label{fig:young}The results of three MD simulations of uniaxial extension of the a-C sample, used to estimate the Young's modulus $Y$. Only a linear parts of the curves at the deformations less than $5\%$ (red) were used to compute $Y$.}
\end{figure}

\begin{figure*}[!]
\includegraphics[width=0.85\textwidth]{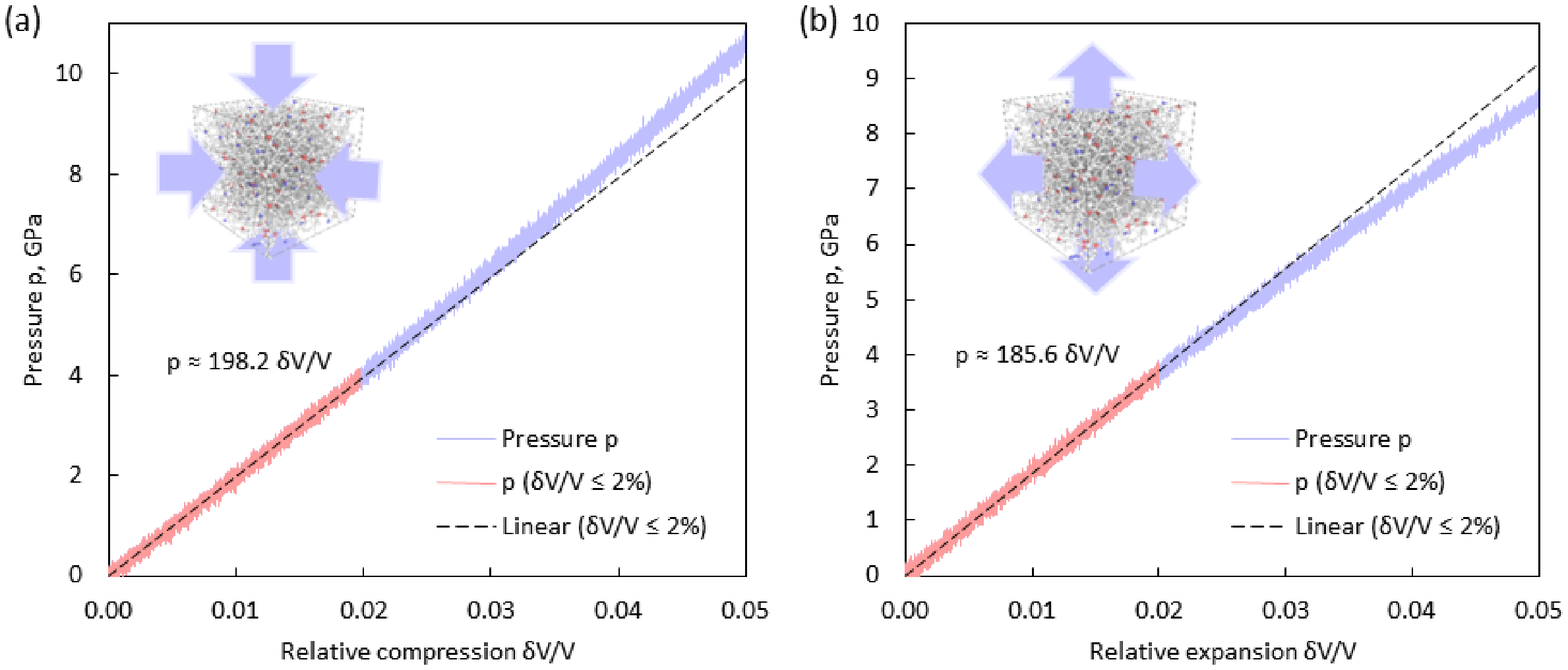}
\caption{\label{fig:bulk} The results of two MD simulations used to estimate the  bulk modulus $K$ during the slow compression (a) and expansion (b). Only the linear parts of the curves at small deformations, less then  $ 2\%$ (red), were exploited. }
\end{figure*}

To estimate Young's $Y$ and the bulk $K$ moduli of amorphous carbon we performed a series of three NEMD simulation of uniaxial loading of a periodic a-C fragment and two simulation of all-round compression and expansion. The initial dimensions of a-C block were $33.4\times33.6\times33.4\ \rm\AA$. The number of atoms was $N=4096$. In the first series, the system underwent uniaxial extension separately along each of the  three axis. The maximal relative deformation used for the estimation of $Y$ was $0.05$, while for $K$  -- up to $0.02$. In both series the deformation rate was rather small,  $\dot\varepsilon=10^7\, {\rm s^{-1}}$; the isothermal condition with $T=300$~K was applied. The obtained stress-strain curves are presented in  Figs.~\ref{fig:young} and \ref{fig:bulk}. To find the moduli $Y$ and $K$ only linear part of the  curves was used. The Young modulus was estimated as the average of three values of $Y$ obtained for separate uniaxial deformations along each of the axis. These values were also used to estimate the standard deviation of $Y$. Similarly, the  bulk modulus  and its error were estimated from the values  of $K$ obtained for the compression and expansion.

As an additional check of the accuracy of our approach we used the experiment-based empirical relation between Young's modulus and the density of amorphous carbon films
\cite{Schneider1997, Schneider1998}, which reads,
$$
\rho_{\rm empir}=1790\left[1 + \frac{Y[{\rm GPa}]}{780} + \left(\frac{Y[{\rm GPa}]}{1620}\right)^2\right][{\rm kg/m^3}].
$$
For the value of $Y=269.2$~GPa, obtained in the present study, this relation gives $\rho\approx 2358\ {\rm kg/m^3}$. It is rather  close to the density of $2178\ {\rm kg/m^3}$ used in our model.


%\newpage
\section{Number of contact atoms and estimate of the bond strength}

\begin{figure}[!]
\includegraphics[width=1.0\columnwidth]{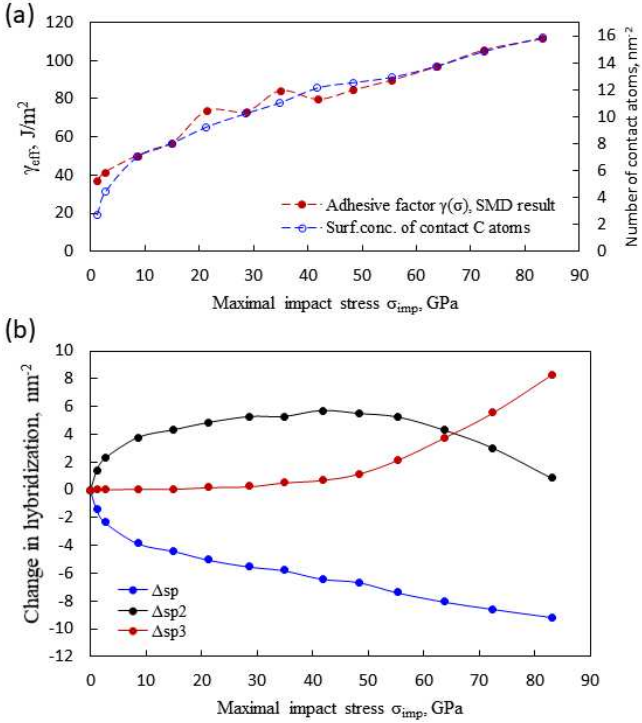}
\caption{\label{fig:contacts} (a) The number of surface $C$ atoms in contact per unit area (blue dots) and the adhesion coefficient (red dots) as the function of the maximal stress $\sigma_{\rm max}$ for the compression of two a-C plates (see the text for detail). The dashed lines are the guides to the eye. (b)  The change of the number of $sp-$, $sp^2-$ and $sp^3$-hybridized surface atoms with increasing maximal stress (see the text for more detail). Lines are the guides to the eye.}
\end{figure}

To find  the number of contacts of carbon atoms between two surfaces $n_{s.c.}$ we  compute the number of $C_1-C_2$ pairs ($C_{1/2}$ is the carbon atom belonging to the first/second surface) with the inter-atomic distance $r_{C_1-C_2} \leq 1.85 {\rm \AA}$, which corresponds to the first  peak of the $C-C$ RDF, see Fig. \ref{fig:rdf}. In Fig.~\ref{fig:contacts}a we compare the dependence of the adhesion coefficient $\gamma$ and the number of $C_1-C_2$ contacts per unit area on the maximal stress $\sigma_{\rm max}$. Apart from two initial points corresponding to the relatively small stress, $n_{s.c.}$ differs from $\gamma$ by a constant factor, corresponding to the average energy of the formed bond $\left< E_{\rm b} \right>$. The number of contact linearly depends on the maximal stress which refers to the linear compression regime of the two surfaces. The smaller value of $n_{s.c.}$ for low $\sigma_{\rm max}$ corresponds to the beginning of the contact, when only part of the surfaces are brought into the contact through the protrudent surface asperities. Fig.~\ref{fig:contacts}b also indicates that for small $\sigma_{\rm max}$ the relative number  of different bonds, $sp$, $sp_2$, and $sp_3$ rapidly varies with the growing stress, and then remains practically constant. This results in a rapid initial decay of the average bond energy $\left< E_{\rm b} \right>$ for low $\sigma_{\rm max}$. The combination of these two factors (rapid increase of $n_{s.c.}$ and decrease of $\left< E_{\rm b} \right>$) for the small values of the maximal stress yields almost linear dependence, Eq. \eqref{eq:gammaeff},  for  $\gamma(\sigma_{\rm max})$ for the whole range of the studied stress.

\begin{figure*}[!]
\includegraphics[width=1.0\textwidth]{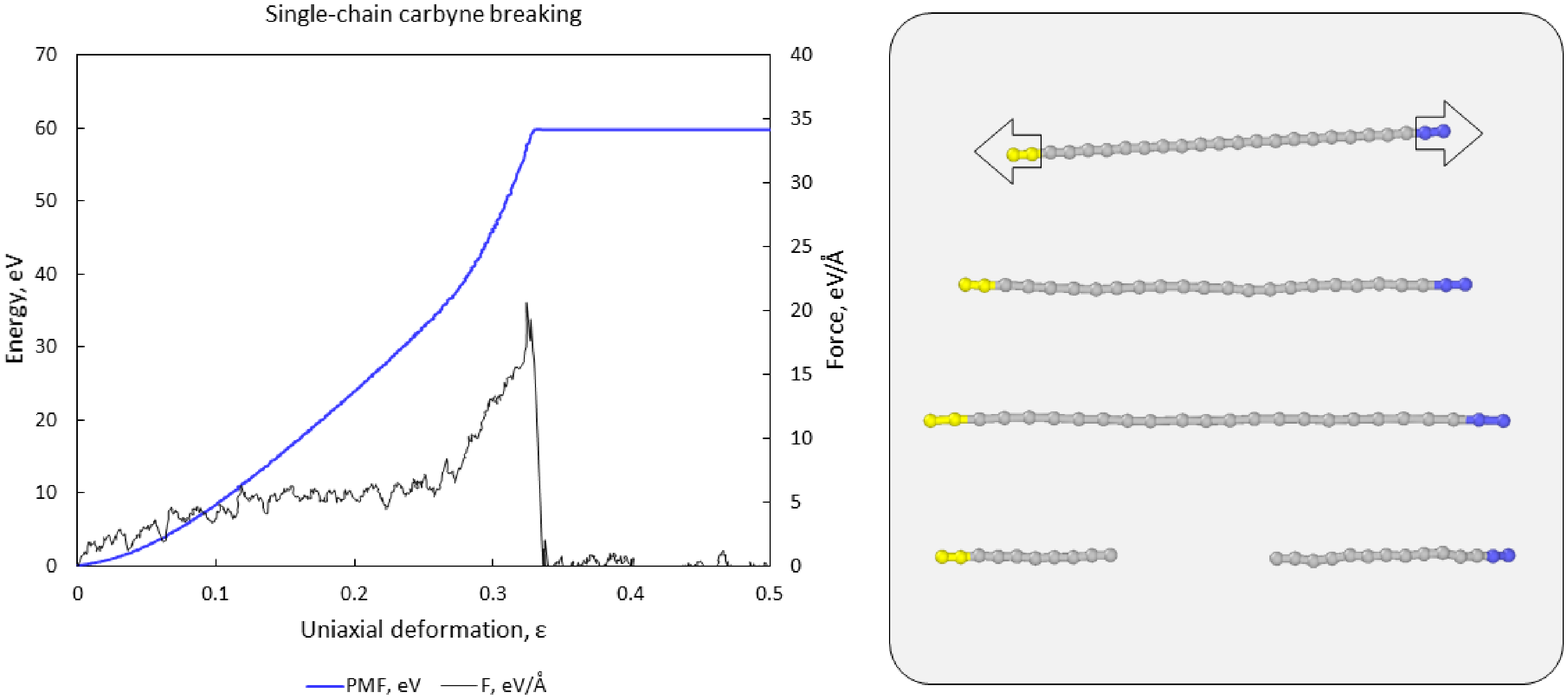}
\caption{\label{fig:carbyne}The results of REBO2-based cv-SMD simulation for the extension of a single carbyne chain. Left panel: The dependence of the potential of the mean force (PMF) (blue line) and of the external force $F$ (gray line) on the uniaxial deformation $\varepsilon$.  Right panel: The series of the subsequent configurations of the carbyne chain during the extension and after the breaking.}
\end{figure*}

We also estimated the work required to break one covalent bond in a single carbyne chain. We performed cv-SMD simulation of extension/breaking of the  chain (see Fig.~\ref{fig:carbyne}). We found that the maximal force before the $sp$-bond between the carbon atom breaks was $F_{\rm max} \approx 20.6\ {\rm eV/\AA}$, (Fig.~\ref{fig:carbyne} the grey line on the left panel). The potential of mean force gives the estimate for the work to break the bond $\Delta {\rm PMF} \approx 59.8$~eV or $9.6\cdot10^{-18}$~J.
Note that the above values characterise the model based on the REBO2 force field; these quantities are needed for the better understanding of background physics. To obtain  accurate estimates  for such characteristics of a real carbyne chain a more complicated model should be used; this is beyond the scope of the current study.

%\nocite{*}

%\nolinenumbers
%\bibliography{agglomeration}
%

\end{document}